\documentclass[prb,notitlepage]{revtex4-1}

\usepackage{amsmath}   
\usepackage{graphicx}   
\usepackage{color}
\usepackage{verbatim}
\usepackage{amssymb}
\usepackage{epsfig}
\usepackage{mathrsfs}  
\usepackage[resetlabels,labeled]{multibib}
\usepackage{censor}

\usepackage[dvipsnames]{xcolor}

\usepackage{cancel}
\usepackage{verbatim}
\usepackage{mathtools}

\begin{document}

\title{Orbital decay in the classroom}

\author{
{Carlos Chaparro$^{1}$, Andrea Ferroglia$^{2,3}$, {Miguel C. N. Fiolhais}$^{1,3,4}$, Luis Gonz\'alez-Urbina$^{1}$, Tomasz Milewski$^{1}$}
\\[3mm]
{\footnotesize {\it 
{$^1$ Science Department, Borough of Manhattan Community College, The City University of New York,}\\ {199 Chambers St, New York, NY 10007, USA} \\ 
{$^2$ Physics Department, New York City College of Technology, The City University of New York, 300 Jay Street, Brooklyn, NY 11201, USA} \\
{$^3$ The Graduate School and University Center, The City University of New York, 365 Fifth Avenue, New York, NY 10016, USA} \\
{$^4$ LIP, Departamento de F\'{\i}sica, Universidade de Coimbra, 3004-516 Coimbra, Portugal}\\
}}
}

\begin{abstract}

The objective of this paper is to provide a pedagogical framework for the phenomenon of orbital decay of satellites in low Earth orbit. The dynamics of orbital decay are derived considering atmospheric drag as the only dissipative mechanism and using an educational approach suitable for undergraduate calculus-based physics and engineering courses. The resulting non-linear first order differential equation for the altitude as a function of time is solved numerically for the isothermal-barotropic atmospheric model with a {fixed} scale height. The model is validated using the uncontrolled reentry data of the Chinese space station Tiangong-1.
\end{abstract}

\maketitle

\section{Introduction}

A common misconception among undergraduate physics students when learning about gravity is to think that there is a nearly pure vacuum just above the K\'arm\'an line, {located  $100$~km above the Earth's surface}, which establishes the boundary between Earth's atmosphere and outer space. {The line introduced by Theodore von K\'arm\'an in 1956 set the edge of space at an altitude above which aerodynamic lift would keep an aircraft aloft without overheating.~\cite{karman}} In fact, satellites, space stations and spacecrafts approaching a planet experience repeated collisions with the upper atmosphere gas molecules. These collisions, {that take place well above the K\'arm\'an line,} lead to loss of mechanical energy, and therefore, a reduction in altitude with time, a phenomenon commonly known as orbital decay. 

Orbital decay is often used to bring down to Earth orbiting objects. While a controlled de-orbit can be very accurate, trajectories for uncontrolled reentries are very complicated, making predictions differ and the possible reentry orbits assorted. For example, in February 2021 NASA landed Perseverance rover in planet Mars’ Jezero crater.~\cite{nasa1} The successful landing was the most accurate to date, a feat for which the trajectory and commands were programmed in advanced due to the 11 minute delay in communication with Earth.~\cite{nasa2} Three months later, however, all eyes on the sky were concerned about the uncontrolled reentry of the 21 metric tons rocket, Chang Zheng 5B (CZ-5B), the core stage of space mission CZ-5B Y2.~\cite{bbc} The different space agencies offered real time predictions for the reentry trajectory and the expected splashdown, but always with large uncertainty margins due to the significant number of degrees of freedom involved in the calculations.~\cite{usatoday} Consequently, the exact time and location of reentry was nearly impossible to predict just before the event took place.

The models that mathematically simulate atmospheric drag in low Earth orbit (from 180 km up to 2,000 km in altitude) take into account parameters such as the density of the atmosphere, sun activity and the shape and the mass of the object. While some complex models~\cite{scalabrin} may also include hypersonic fluid physics, thermal radiation, heat and mass transfer, and shock waves, other simpler approaches consider only an atmospheric model to predict the atmosphere density.~\cite{kennewell} {For example, Knipp et al.~\cite{knipp} developed a spreadsheets-based simulation-laboratory to produce satellite orbits calculations based on different atmospheric models.} In this simplified framework, the trajectory of an object in low Earth orbit can be inferred from the loss of mechanical energy due to the dissipative atmospheric drag force.~\cite{marcos} As the satellite’s orbit decays, the density of the atmosphere increases with the lower altitude, and therefore the rate of decay increases producing a self-reinforcing feedback effect.

Using classical mechanics and considering only the effect of atmospheric drag, orbital decay can be predicted with high accuracy down to an altitude of 180~km.~\cite{kennewell} For a satellite in circular orbit, \emph{i.e.} small eccentricity, the orbit can be considered to remain approximately circular throughout the decaying process up to this altitude. Above an altitude of 180~km, the impact of the atmospheric drag is small enough that the gravitational force may stay approximately perpendicular to the velocity vector while the satellite spirals down to Earth over a long period of time - orbital decay may take several months while a single orbit at that altitude takes approximately ninety minutes to complete.


The objective of this work is to study the phenomenon of orbital decay of satellites in circular low Earth orbit in the context of an undergraduate calculus-based physics course. The dynamics of orbital decay are derived considering atmospheric drag as the only dissipative mechanism. The resulting non-linear first order differential equation for the altitude as a function of time is solved numerically for the isothermal-barotropic atmospheric model. Results are presented for the orbital decay data of the Chinese space station Tiangong-1 in 2018.~\cite{pardini,vellutini} The calculations in the work were also performed by undergraduate students within the scope of a calculus-level university physics course at the Borough of Manhattan Community College (BMCC) of the City University of New York (CUNY), in order to show that this approach can be used as an approximation activity for calculus-based physics classes.



\section{Dynamics of Orbital Decay}

\begin{figure}[h]
    \centering
\includegraphics[height=0.40\textwidth]{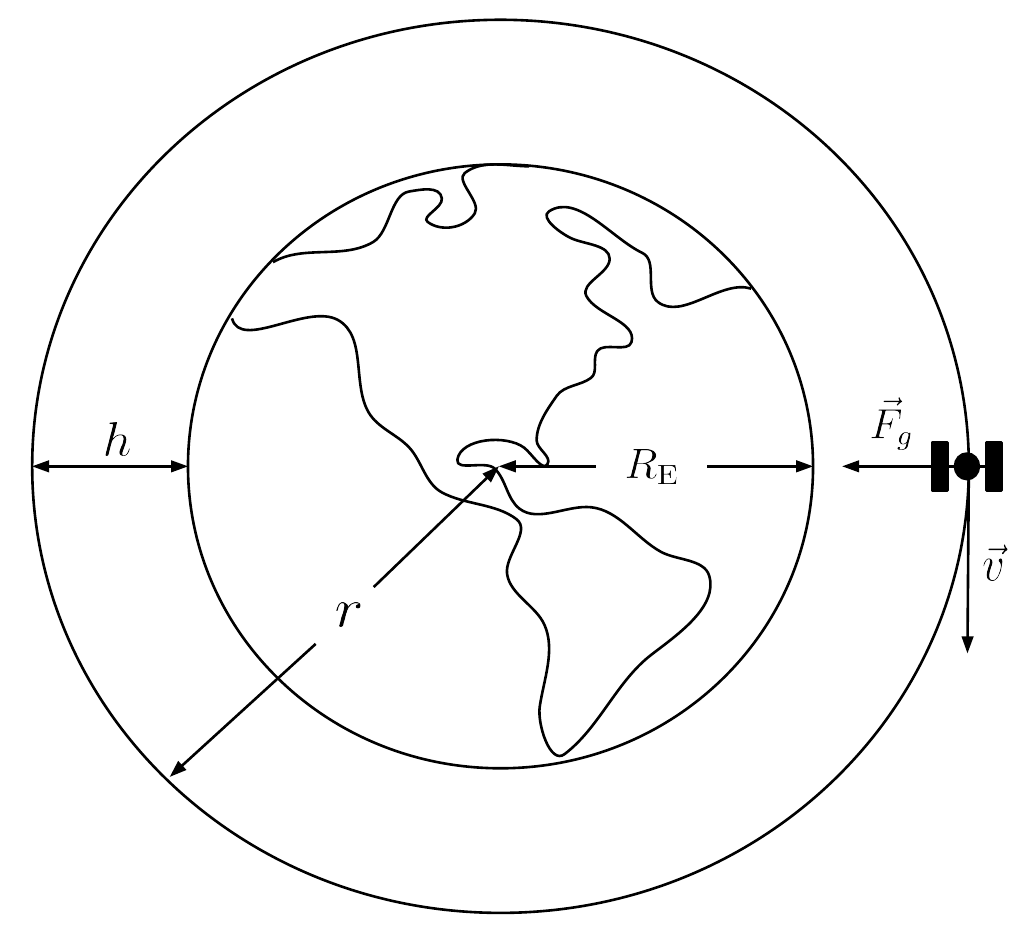}
    \caption{Schematic representation (not to scale) of a satellite in circular orbit around planet Earth without atmospheric drag. The radius of the Earth and the altitude of satellite are  represented by $R_\textrm{E}$ and $h$, respectively.}
    \label{fig:orbitaldecay1}
\end{figure}

Before studying the case of orbital decay, one shall consider the case of a circular orbit without any form of mechanical energy dissipation, as shown in Figure~\ref{fig:orbitaldecay1}. The force of gravity on the satellite, $\vec{F}_g$, acts as a centripetal force, so that, by applying Newton's second law and the relation between centripetal acceleration and the satellite's linear velocity, the speed of the satellite can be expressed  as,
\begin{equation}
 v = \sqrt{\frac{GM}{r}} \, ,
 \label{eq:velocity}
\end{equation}
where $G$ is the gravitational constant, $M$ is the mass of Earth, and $r$ is the radius of the trajectory. As a result of Eq.~(\ref{eq:velocity}), the conserved mechanical energy of the system is
\begin{equation}
 E =  -\frac{GMm}{r} + \frac{1}{2}mv^2 = -\frac{GMm}{2r} \, ,
 \label{eq:energy}
\end{equation}
where $m$ is the mass of the satellite. {The values of the gravitational constant, mass of planet Earth, mass of Tiangong-1 and radius of the Earth used in this study are presented in Table~\ref{tab:parameters}.}

\begin{table}[h]
\centering
 \begin{tabular}{c c c c}
 \hline
 $G$ & $M$ & $m$ & $R_{\textrm{E}}$ \\ [0.5ex]
 \hline
  \,\,\,$6.674\times10^{-11}$~N$\cdot$m$^2\cdot$kg$^{-2}$ \,\,\, & \,\,\,$5.972\times10^{24}$~kg\,\,\, & \,\,\,$8.506\times10^3$~kg\,\,\, & \,\,\,$6.378\times 10^6$~m\,\,\, \\
  \hline
  \end{tabular}
  \caption{{Values of the gravitational constant, mass of planet Earth, mass of Tiangong-1 and radius of the Earth, respectively.}}
   \label{tab:parameters}
\end{table}

\begin{figure}
    \centering
\includegraphics[height=0.40\textwidth]{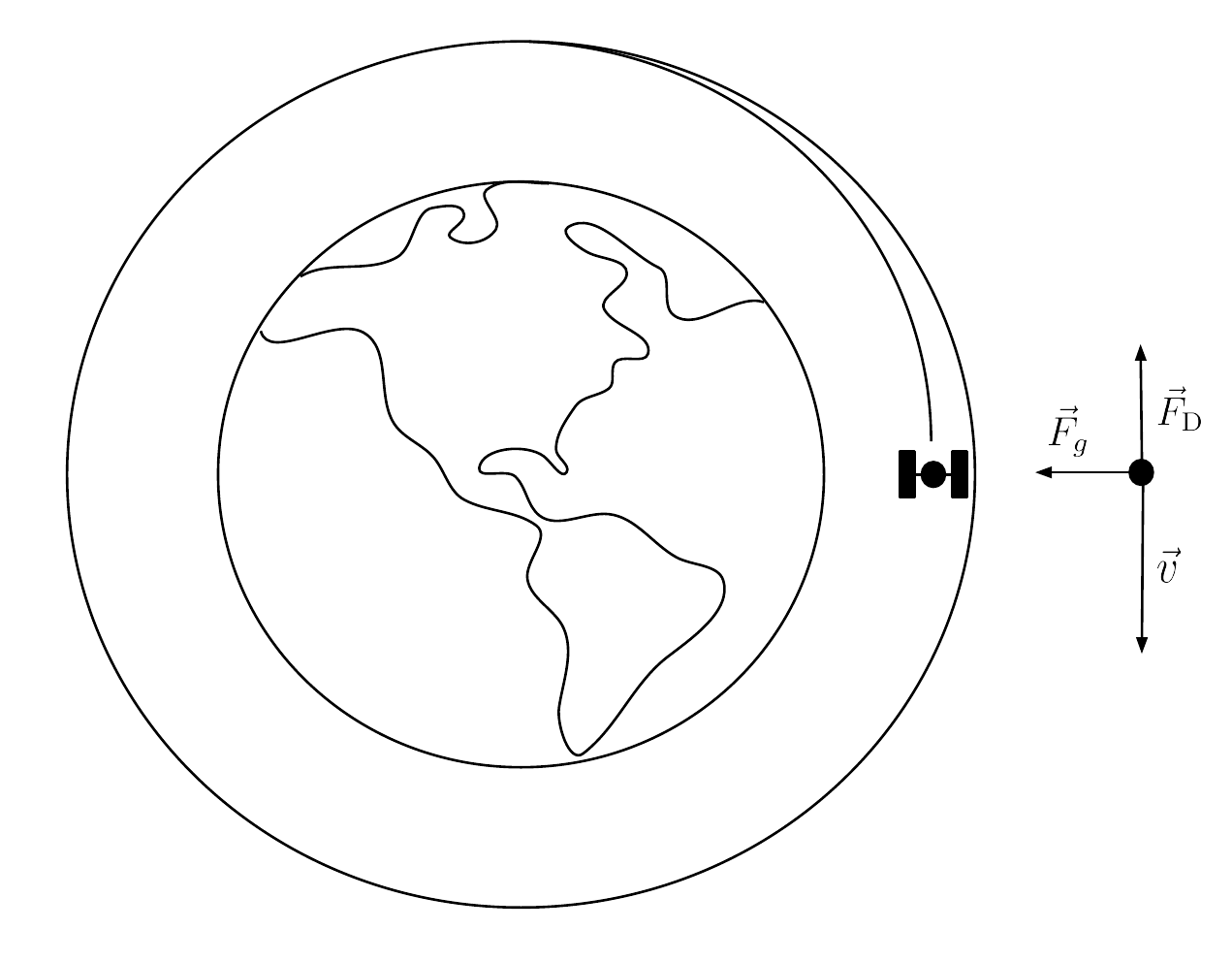}
    \caption{{Schematic representation (not to scale) of a satellite orbiting planet Earth in orbital decay. The figure shows a representation of the decrease in altitude due to atmospheric drag as well as the gravitational and drag forces applied on the satellite.}}
    \label{fig:orbitaldecay2}
\end{figure}

If the satellite is at a lower circular orbit, the drag force must be considered; {this force  points  in the direction  opposite to the velocity vector, as shown in Figure~\ref{fig:orbitaldecay2}.} The presence of this non-conservative force leads to the loss of mechanical energy over time, and therefore {to} a decrease in altitude. The dissipated power due to drag can be written as $\mathscr{P} = \vec{F}_D \cdot \vec{v}$, where $\vec{F}_D$ is the drag force, and the equation for the time derivative of the satellite's mechanical energy becomes,
\begin{equation}
    \frac{\textrm{d}}{\textrm{d}t}\left( -\frac{GMm}{2r}\right) = \vec{F}_D \cdot \vec{v}  \, .
\label{eq:power}
\end{equation}
The magnitude of the drag force can be obtained from the drag equation,
\begin{equation}
 F_{\textrm{D}} = \frac{1}{2} \rho(r) C_\textrm{D} A v^2 \, ,
\label{eq:drag}
\end{equation}
where $\rho(r)$ is the atmospheric density, $C_{\textrm{D}}$ is the dimensionless drag coefficient, $A$ is the satellite's cross sectional area and $v$ is the satellite's speed relative to the air. It should be observed that, because of the size and speed of the objects involved, the atmospheric drag force is quadratic in the velocity. Consequently, the drag force cannot be described by Stoke's law, a law often introduced in university physics courses, which is linear in the velocity and  which describes the drag force in the case of  small objects moving at low speed, such as for example oil drops in Millikan's experiment to measure the electron charge.

As previously mentioned, if the timescale of orbital decay is assumed to be much longer than the period of one orbit, it is reasonable to consider that the trajectory of the satellite remains approximately circular as the altitude decreases. Consequently, the substitution of Eqs. (\ref{eq:velocity}) and (\ref{eq:drag}) in Eq.~(\ref{eq:power}) leads to,
\begin{equation}
     \frac{GMm}{2} \frac{\dot{r}}{r^2}  =  - \frac{1}{2} \rho(r) C_{\textrm{D}} A \left ( \frac{GM}{r} \right )^{3/2} \, .
\end{equation}
The simplification of this expression leads to a non-linear first order differential equation for the radius of the satellite's orbit as a function of time,
\begin{eqnarray}
    \dot{r}  & = & - k \sqrt{r} \rho(r) \, ,
\label{eq:radius}
\end{eqnarray}
where $ k = \sqrt{GM} A_{\textrm{eff}}/m$ and $A_{\textrm{eff}}=C_{\textrm{D}} A$ is the effective cross-sectional area. The same equation can be written in terms of the satellite's altitude $h$ above Earth's surface,
\begin{equation}
    \dot{h}   =  - k \sqrt{R_\textrm{E}+h} \, \rho(h) \, ,
\label{eq:altitude}
\end{equation}
where $R_\textrm{E}$ is the radius of the Earth and $r= R_\textrm{E}+h$.

In order to determine the time evolution of the satellite's altitude, it is necessary to use a distribution of atmospheric density as input. {The density of the atmosphere can be described in first approximation as an isothermal-barotropic atmospheric model with a fixed scale height.  In this model, the temperature is assumed constant and the pressure is a function of the density only. The density as a function of altitude can be easily obtained from the condition of hydrostatic equilibrium and the ideal gas law,~\cite{isothermal}
\begin{equation}
    \rho (h)  =  \rho_0 \exp\left( - \frac{h-h_0}{H} \right) \, ,
\end{equation}
where $\rho_0$ is the density at $h = h_0$ and $H$ is the scale height. The scale height is defined as,
\begin{equation}
    H  =  \frac{RT}{M_{\textrm{air}} g} \, ,
\end{equation}
where $R$ is  the ideal gas constant, $T$ is the temperature, $M_{\textrm{air}}$ is the molar mass of air and $g$ is the acceleration of gravity. In this model, the atmospheric density drops by a factor $e$ for every increase in altitude $H$. For this particular study, the following  equation for the density is used in the numerical calculation the orbital decay of Tiangong-1, 
\begin{equation}
    \rho (h)  =  6 \times 10^{-10} \exp\left( - \frac{h-175~\textrm{km}}{H} \right)\, \textrm{kg}/\textrm{m}^3 \, .
\end{equation}
This equation, adapted from the Australian Space Weather Agency’s model for an isothermal-barotropic model,~\cite{kennewell} is calibrated so the density is $6 \times 10^{-10}$~kg/m$^3$ at an altitude of 175~km. The model is accurate up to an altitude of 500~km.}

{\section{Tiangong-1 orbital reentry data and student activity}}

\begin{figure}[h]
    \centering
\includegraphics[height=0.4\textwidth]{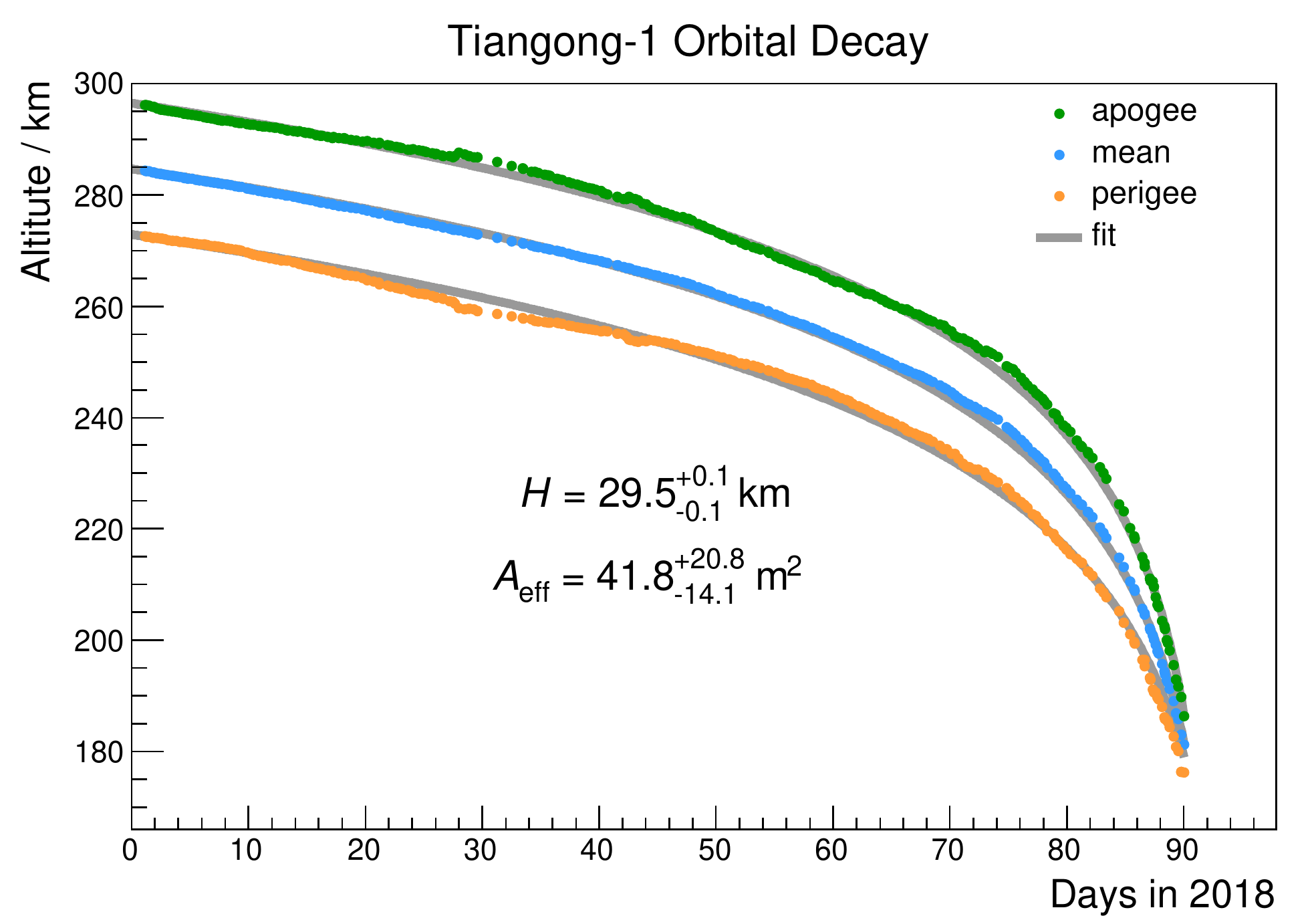}
    \caption{{Tiangong-1 orbital decay data and best fit line. The plot shows the recorded daily data for the altitude of the apogee (green) and perigee (orange), and mean altitude (blue). The calculated fit lines for the apogee, perigee and mean altitude are represented in gray.}}
    \label{fig:orbitaldecay3}
\end{figure}

A few months after losing contact with ground control in March 2016, Tiangong-1's orbit started a slow decay until it finally reentered the atmosphere on 2 April 2018 at 00:16 UTC.~\cite{UN1} The orbital data for Tiangong-1 during 2018 is presented in Figure~\ref{fig:orbitaldecay3}.~\cite{pardini} {The plot shows the evolution of the altitude of the apogee (green) and perigee (orange) as a function of the days in 2018, as well as the mean altitude of the space station (blue). The apogee, mean, and perigee altitudes are also compared to the fit lines (gray) determined using the isothermal-barotropic atmospheric model with a fixed scale height presented in the previous section.~\cite{kennewell} The theoretical lines were calculated by numerically solving the differential Eq.~(\ref{eq:altitude}) with the parameters presented in Table~\ref{tab:parameters}, using the \texttt{NDSolve} function in \texttt{Mathematica}.~\cite{mathematica}}

{The result was fitted to the mean altitude data using two free parameters: the fixed scale height $H = 29.5$~km and the effective cross-section $A_{\textrm{eff}} = 41.8$~m$^2$. A fit was also performed to the apogee and perigee data -- the fit value for the fixed scale height is the same as for the mean altitude, while the obtained effective cross-sections are $62.6$~m$^2$ and $27.7$~m$^2$ for the apogee and perigee, respectively. The variation in these values can be used as an uncertainty in the result for the effective cross-section obtained for the mean altitude, yielding $A_{\textrm{eff}} = 41.8^{+20.8}_{-14.1}$~m$^2$. {The fit line is extremely sensitive to the value of the scale height, \emph{i.e.} a small variation creates a significant disparity between the data and the fit lines in all three cases. Therefore, an uncertainty of 0.1~km was included in the scale height as a conservative estimate of the uncertainty: A variation of 0.1~km on the scale uncertainty produces a change of 5-10\% on the predicted altitude for the last days of the orbital decay.} It is worth noting that the result for the effective cross-section is physically appropriate given the size of 10 meters by 3 meters of Tiangong-1. Similarly, a scale height of approximately 30~km in the upper atmosphere is also a reasonable result.}

{The value of the fit parameters was obtained via trial and error during the classroom activity, instead of a $R^2$ method, due to the large number of data points. This can be circumvented during the activity by comparing the different results obtained by students and narrowing down the sets of parameters that best adjust to the data.} This calculation also assumed that Tiangong-1 followed approximately successive circular trajectories throughout orbital decay, which is a good approximation given that {the differences between the apogee and perigee never exceeded 0.3\% of the mean trajectory radius.} {Moreover, at an altitude of $280$~km, the speed of a satellite is approximately 7.7~km/s, completing a circular orbit every 90~minutes -- a total of 16 orbits per day. As such, dozens or hundreds of orbits are necessary to achieve a significant decrease in altitude.} The remarkable agreement between the calculated fit line and the mean altitude data is visible in Figure~\ref{fig:orbitaldecay3}.
{It is also possible to obtain similar results with the Australian Space Weather Agency isothermal-barotropic atmospheric model with a variable scale height~\cite{kennewell}. This model is particularly useful for higher altitudes (up to 500~km) than the case studied in this activity, for which the variable scale height provides a significant correction.}

{The activity was performed online via Zoom due to the COVID-19 pandemic with calculus-based University Physics I and II students at BMCC. After a general introduction to the topic for half an hour, students were given the task of deriving Eq.~(\ref{eq:altitude}). At this point groups of two to fours students were formed using breakout rooms on Zoom, with at least one A-student per group to provide support to those in need. Most groups were able to reach the desired result, however, the calculation was repeated with the class as a whole in order to solidify concepts. The equation was then implemented in \texttt{Mathematica} to be solved numerically using \texttt{NDSolve} for different values of the effective cross-section and temperature parameters to be fitted to the data points. As most students had never used \texttt{Mathematica} before, this was done step-by-step so they could recreate the code lines in their own computers. After several trial-and-error attempts made by students, the class agreed on the fit parameter values presented in Figure~\ref{fig:orbitaldecay3}. The discussion followed on the validity of the model, and if it could be used to make valid predictions. This resonated with the students due to the broadly publicized case of the uncontrolled reentry of CZ-5B.}

It should be noted that it is also possible to extend this activity to obtain a graphical representation of the spiraling orbital decay. By combining Eq.~(\ref{eq:radius}) and the fact that $\delta\theta / \delta t \approx v / r$, it is possible to calculate a parametric plot for $r(\theta)$. Even though this is beyond the scope of this work, mainly aimed at undergraduate science and engineering students at a community college, this extended activity could potentially be a good fit for junior or senior physics majors attending an analytical mechanics or computational physics course. \\

\section{Conclusions}

Students in a calculus-based University Physics courses (PHY215 and PHY225) at BMCC performed the derivation of the differential equation for the altitude of a satellite in orbital decay, assuming approximately circular orbits. By choosing different combinations of temperature and effective cross-section parameters, students calculated the numerical solution of the differential equation, and {the} corresponding fit line that adjusted to the data for the orbital decay of Tiangong-1 up to a mean altitude of 180~km. The results led to an impressive agreement between the recorded data and the fit line, showing the validity of the circular orbit approximation and the isothermal-barotropic atmospheric model with a {fixed scale} height.

The activity was particularly successful in solidifying concepts of Universal Gravitation, non-conservative forces and first order differential equations among students. While the numerical solution for the differential equation was obtained in \texttt{Mathematica} in this exercise, alternatively students can also be asked to code directly a numerical method in \texttt{MATLAB} or \texttt{Python}. This is especially useful in an applied computational physics course.

\acknowledgements

{The authors would like to thank Dr. Luciano Anselmo and Dr. Carmen Pardini for providing the data used in this}\\ {paper for the orbital decay of Tiangong-1.}

\end{document}